\documentclass[10pt,conference]{IEEEtran}
\usepackage{amssymb,amsmath,amsfonts,bm}
\usepackage{graphicx}

\begin{document}

\title{Reducing the Error Floor}

\author{
\authorblockN{Michael Chertkov}
\authorblockA{ Center for Nonlinear Studies \& Theory Division,
Los Alamos National Laboratory, Los Alamos, NM 87545
}}

\maketitle

\begin{abstract}
\footnote{Invited talk at ITW '07 workshop, Lake Tahoe, Sep 2-5,
2007.} We discuss how the loop calculus approach of [Chertkov,
Chernyak '06], enhanced by the pseudo-codeword search algorithm of
[Chertkov, Stepanov '06] and the facet-guessing idea from [Dimakis,
Wainwright '06], improves decoding of graph based codes in the
error-floor domain. The utility of the new, Linear Programming
based, decoding is demonstrated via analysis and simulations of the
model $[155,64,20]$ code.
\end{abstract}

\section{Introduction}

A new era has begun in coding theory with the discovery of graphical
codes -- low-density parity check codes (LDPC)
\cite{63Gal,95WLK,99Mac,03RU} and turbo codes\cite{93BGT}. These
codes are special, not only because they can virtually achieve the
error-free Shannon limit, but mainly because a family of
computationally efficient approximate decoding schemes is readily
available. This family includes iterative Belief Propagation (BP),
or simply message-passing, decoding \cite{63Gal,95WLK,99Mac} and
Linear Programming (LP) decoding \cite{05FWK}.

When operating at moderate noise values these decoding algorithms
show performance comparable to the ideal, but computationally not
feasible, Maximum Likelihood (ML) and Maximum-a-Posteriori
decodings. However sub-optimality of the approximate decodings
becomes a handicap at large SNR, in the so-called error-floor regime
\cite{03Ric}. An error-floor typically emerges due to the low-weight
fractional pseudo-codewords \cite{01FKKR,03KV,03Ric,05VK,Pascal}, or
if one uses a noise space description, due to the instantons
\cite{05SCCV} -- the most probable erroneous configurations of the
noise. Much effort has been invested in recent years on
understanding the pseudo-codewords/instantons and thus  the error
floor behavior of the graphical codes. Also, there were a few
attempts at decoding improvement. The Facet-Guessing (FG) algorithm
of \cite{06DW} and the Loop Erasure algorithm of \cite{06CCc}
constitute two recent advances in this direction.

Let us review these {\bf relevant prior results:}

\noindent $\bullet$ LP decoding can be considered as an asymptotic,
large SNR, version of the BP decoding \cite{03WJ,04VK} where the
latter is understood as the absolute minimum of the Bethe free
energy \cite{05YFW}. The LP decoding minimizes the linear functional
of beliefs, called the energy functional, under a set of non-strict
inequality and equality constraints \cite{05FWK}. (One can also
consider a small polytope formulation of the problem, where all
constraints are non-strict inequalities.)

\noindent $\bullet$ The error-floor of an LDPC code is typically due
to dangerous low-weight pseudo-codewords which are fractional, i.e.
which are not codewords \cite{03Ric}.

\noindent $\bullet$ The dangerous pseudo-codewords are rare and the
pseudo-codeword search algorithm \cite{06CS} is an efficient
heuristic for finding the troublemakers. This algorithm is based on
LP decoding.

\noindent $\bullet$ The Loop Calculus of \cite{06CCa,06CCb}
introduces the Loop Series, which is an explicit finite expression
for the MAP decoding partition function in terms of loop
contributions defined on the graphical representation of the
respective inference/decoding problem. Each loop contribution is
calculated explicitly from a solution of the BP equations and it is
represented as a product of terms along the loop. The LP limit of
the Loop Calculus is well defined.

\noindent $\bullet$ One experimentally verifies, and otherwise
conjectures, that all the dangerous pseudo-codewords are explained
in terms of a small number of critical loops \cite{06CCc}.
Typically, the respective partition function can be well
approximated in terms of the bare LP/BP term and one critical term
associated with a single connected critical loop. The bare term and
the critical terms are comparable while all other terms in the Loop
Series are much smaller. An efficient heuristic algorithm for
finding the critical loop has been constructed. It is based on
representing the loop contribution as a product of terms along the
loop, each smaller than or equal to unity in absolute value, and
pre-selecting elements of the critical loop to be larger than a
threshold close to unity.

\noindent $\bullet$ BP, corrected by accounting for the critical
loop, and its simplified LP version, coined the LP-erasure
\cite{06CCc}, are algorithms improving BP/LP. These algorithms,
applied when LP/BP fails, consist of modifying BP/LP along the
critical loop. Thus LP-erasure modifies log-likelihoods everywhere
along the critical loop (lowering log-likelihoods in absolute
value). It was shown experimentally (on the example of the test
$[155,64,20]$ code introduced in \cite{01TSF}) that the LP-erasure
is capable of correcting all the dangerous fractal pseudo-codewords,
previously found with the pseudo-codeword search algorithm
\cite{06CS,07CSa}.

\noindent $\bullet$ The Facet Guessing (FG) algorithm \cite{06DW} is
a graph local improvement of the LP algorithm. It applies if LP
decoding does not succeed. A non-active facet/inequality, i.e. the
one with the vertex/solution lying in the strict inequality domain,
is selected. The original LP problem is modified so that the
selected facet is enforced to be in its active (equality) state. The
number of non-active facets for a dangerous fractional
pseudo-codeword is typically small. (It is provably small for the
expander codes.) One constructs a full set of the single-facet
modified LP problems (where the number of problems is thus equal to
the number of active facets) or otherwise selects a random subset of
the LP problems. Running LP decoding for the set of the modified
problems, one chooses solution with the lowest energy functional and
calls it the outcome of the facet guessing algorithm. It was shown
experimentally in \cite{06DW} that the facet guessing algorithm
improves the bare LP decoding.

{\bf Results reported in this paper:}

\noindent $\bullet$ We introduce the Bit Guessing (BG) algorithm,
which is a simplified version of the Facet Guessing algorithm of
\cite{06DW}, enforcing single bits to be in $0$ or $1$ state. We
apply the algorithm to the set of pseudo-codewords found for the
bare LP decoding of the test $[155,64,20]$ code. It was found that
all the fractional dangerous pseudo-codewords are corrected by the
BG algorithm.

 \noindent $\bullet$ For each of the
dangerous pseudo-codewords, we identify the set of bits where local
BG leads to correct decoding and compare this set with the set of
bits forming a critical loop found via the critical loop search
algorithm of \cite{06CCc}.  The comparison shows very strong
correlations between the two sets: fixing a bit from the critical
loop almost always leads to correct decoding. This suggests that,
since the critical loop is relatively small, it is advantageous to
use the loop series and the critical loop analysis to pre-select the
set of single-bit corrected LP schemes in the BG algorithm. (One
will only need to consider fixing bits along the critical loop.)
Moreover, we consider the Loop Guided Guessing (LGG) algorithm built
on top of the bare LP with only one or two modified LP runs. The
modified LP scheme is constructed by adding to the bare LP scheme an
equality fixing the value of a randomly selected bit from the
critical loop.

 \noindent $\bullet$ We test the
LGG algorithm on the set of the LP-erroneous configurations of the
$[155,64,20]$ code for the Additive White Gaussian Noise (AWGN)
channel found in the finite Signal-to-Noise Ratios (SNR) Monte Carlo
simulations of \cite{07CSa}. We show that the LGG algorithm greatly
improves the bare LP algorithm and it also performs significantly
better than the LP-erasure algorithm of \cite{06CCc}.

\section{Error-Floor analysis of BP/LP decoding: Brief Review of Prior Results}

\subsection{Belief Propagation and Linear Programming}

We consider a generic linear code, described by its parity check
$N\times M$ sparse matrix, $\hat{H}$, representing $N$ bits and $M$
checks. The codewords are these configurations,
${\bm\sigma}=\{\sigma_i=0,1| i=1,\dots,N\}$, which satisfy all the
check constraints: $\forall \alpha=1,\dots,M$, $\sum_i H_{\alpha
i}\sigma_i=0$ (mod~$2$). A codeword sent to the channel is polluted
and the task of decoding becomes to restore the most probable
pre-image of the output sequence,  ${\bm x}=\{x_i\}$. The
probability for ${\bm\sigma}$ to be a pre-image of ${\bm x}$ is
\begin{eqnarray}
{\cal P}({\bm\sigma}|{\bm x}) &=& p({\bm\sigma}|{\bm x})Z^{-1},\quad
Z=\sum_{\bm\sigma}p({\bm\sigma}|{\bm x}),\label{Psx1}\\
p({\bm\sigma}|{\bm x}) &=& \prod_\alpha
\delta\biggl(\prod_{i\in\alpha}(-1)^{\sigma_i},1\biggr)
\exp\biggl(-\sum_ih_i\sigma_i\biggr), \label{Psx}
\end{eqnarray}
where one writes $i\in\alpha$ if $H_{\alpha i}=1$; $Z$ is the
normalization coefficient (so-called partition function); the
Kronecker symbol, $\delta(x,y)$, is unity if $x=y$ and it is zero
otherwise; and ${\bm h}$ is the vector of log-likelihoods dependent
on the output vector ${\bm x}$. In the case of the AWGN channel with
the {\rm SNR} ratio, $E_c/N_0=2s^2$, the bit transition probability
is $\sim \exp(-2s^2(x_i-\sigma_i)^2)$, and the log-likelihood
becomes $h_i=s^2(1-2x_i)$. The optimal block-MAP (Maximum
Likelihood) decoding maximizes ${\cal P}({\bm\sigma}|{\bm x})$ over
${\bm\sigma}$, $ \arg\max_{\bm\sigma}{\cal P}({\bm\sigma}|{\bm x})$
and symbol-MAP operates similarly, however in terms of the marginal
probability at a bit
$\arg\max_{\sigma_i}\sum_{{\bm\sigma}\setminus\sigma_i}{\cal
P}({\bm\sigma}|{\bm x})$. One can also think formally about ML in
terms of MAP,  i.e. in terms of summation over all possible
configurations of ${\bm\sigma}$ however with the weight and the
partition function in Eqs.~(\ref{Psx1},\ref{Psx}) transformed
according to
\begin{equation}
\ln( p({\bm\sigma}|{\bm x}))\to \rho\ln( p({\bm\sigma}|{\bm
x})),\quad \rho\to+\infty. \label{rho}
\end{equation}

BP and LP decodings should be considered as computationally
efficient but suboptimal substitutions for MAP and ML. Both BP and
LP decodings can be conveniently derived from the so-called
Bethe-Free energy approach of \cite{05YFW} which is briefly reviewed
below. In this approach trial probability distributions, called
beliefs, are introduced both for bits and checks, $b_i$ and
$b_\alpha$, respectively. The set of bit-beliefs, $b_i(\sigma_i)$,
satisfy equality and inequality constraints that allow convenient
reformulation in terms of a bigger set of beliefs defined on checks,
$b_\alpha({\bm\sigma}_\alpha)$, where, ${\bm
\sigma}_\alpha=\{\sigma_i|i\in\alpha,\sum_i H_{\alpha
i}\sigma_i=0\mbox{ (mod~$2$)}\}$, is a local codeword associated
with the check $\alpha$. The equality constraints are of two types,
normalization constraints (beliefs, as probabilities, should sum to
one) and compatibility constraints:
\begin{equation}
\forall i,\ \forall\alpha\ni i:\
b_i(\sigma_i)=\sum\limits_{\sigma_\alpha\setminus\sigma_i}b_\alpha({\bm
\sigma}_\alpha),\
\sum\limits_{{\bm\sigma}_\alpha}b_\alpha({\bm\sigma}_\alpha)=1.\label{comp}
\end{equation}
Additionally, all the beliefs should be non-negative and smaller
than or equal to unity. The Bethe Free energy is defined as the
difference of the self-energy and the entropy, $F=E-S$:
\begin{eqnarray}
 && \!\!\!\!\!\!\!\!\!\!\!\!\!\!\!\!\!\!\!\!\!
 E\!=\sum_ih_i\sum_{\sigma_i}\!\sigma_ib_i(\sigma_i)\quad\mbox{and}\label{E_Bethe}\\
 && \!\!\!\!\!\!\!\!\!\!\!\!\!\!\!\!\!\!\!\!\!
  S\!\!=\!-\!\!\sum\limits_\alpha\! \sum_{{\bm \sigma}_\alpha}\!
  b_\alpha\!({\bm\sigma}_\alpha\!)
 \!\ln b_\alpha\!({\bm\sigma}_\alpha\!)\!
 +\!\!\sum\limits_i\!\sum\limits_{\sigma_i}
 (q_i\!-\!1)b_i(\sigma_i)\!\ln\! b_i(\sigma_i).
 \label{S_Bethe}
\end{eqnarray}
Optimal configurations of beliefs minimize the Bethe Free energy
subject to the equality constraints (\ref{comp}). Introducing the
constraints as the Lagrange multiplier terms to the effective
Lagrangian and looking for the extremum with respect to all possible
beliefs leads to
\begin{eqnarray}
 &&\!\!\!\!\!\!\!\!\!\! b_\alpha({\bm
 \sigma}_\alpha)=\frac{\exp\left(\sum_{i\in\alpha}(h_i/q_i+\eta_{\alpha
 i})(1-2\sigma_i)\right)}{\sum_{{\bm\sigma}_\alpha}\exp\left(\sum_{i\in\alpha}(h_i/q_i+\eta_{\alpha
 i})(1-2\sigma_i)\right)},\label{ba}\\
 &&\!\!\!\!\!\!\!\!\!\! b_i(\sigma_i)=\frac{\exp\left((\eta_{\alpha
 i}+\eta_{i\alpha})(1-2\sigma_i)\right)}{2\cosh\left(\eta_{i\alpha}+\eta_{\alpha
 i}\right)},\label{bi}
\end{eqnarray}
where the set of $\eta$ fields (which are Lagrange multipliers for
the compatibility constraints) satisfy
\begin{eqnarray}
 \eta_{\alpha i}=h_i+\sum_{\beta\ni i}^{\beta\neq \alpha}\eta_{i\alpha},\ \
 \eta_{i\alpha}=\tanh^{-1}\left(\prod_{j\in\alpha}^{j\neq
 i}\tanh\eta_{\alpha j}\right).
 \label{eta}
\end{eqnarray}
These are the BP equations for LDPC codes written in its standard
form. These equations are often described in the coding theory
literature as stationary point equations for the BP (also called the
sum product) algorithm and then $\eta$ variables are called
messages. The BP algorithm, initialized with $\eta_{i\alpha}=0$,
solves Eqs.~(\ref{eta}) iterating it sequentially from right to
left. Possible lack of the iterative algorithm convergence (to the
respective solution of the BP equation) is a particular concern, and
some relaxation methods were discussed to deal with this problem
\cite{06SCb}.

LP is a close relative of BP which does not have this unpleasant
problem with convergence. Originally, LP decoding was introduced as
a relaxation of ML decoding \cite{05FWK}. It can thus be restated as
$ \arg\min_{{\bm\sigma}\in {\cal
P}}\biggl(\sum_ih_i\sigma_i\biggr)$, where ${\cal P}$ is the
polytope spanned by all the codewords of the code. Looking for
${\bm\sigma}$ in terms of a linear combination of the codewords,
${\bm \sigma}_v$: ${\bm\sigma}=\sum_v\lambda_v{\bm \sigma}_v$, where
$\lambda_v\geq 0$ and $\sum_v\lambda_v=1$, one observes that the
block-MAP turns into a linear optimization problem. The LP-decoding
algorithm of \cite{05FWK} proposes to relax the polytope, expressing
${\bm\sigma}$ in terms of a linear combination of local codewords
associated with checks, ${\bm\sigma}_\alpha$. We will not give
details of this original formulation of LP here because we prefer an
equivalent formulation, elucidating the connection to BP decoding.
One finds that the BP decoding, understood as an algorithm searching
for a stationary point of the BP equations, turns into LP decoding
in the asymptotic limit of large {\rm SNR}. Indeed in this special
limit, the entropy terms in the Bethe free energy can be neglected
and the problem turns into minimization of a linear functional with
a set of linear constraints. The relation between BP and LP was
noticed  in \cite{03WJ,03KV} and it was also discussed in
\cite{04VK,06CCc}. Stated in terms of beliefs, LP decoding minimizes
the self-energy part (\ref{E_Bethe}) of the full Bethe Free energy
functional under the set of linear equality constraints (\ref{comp})
and also linear inequalities guaranteeing that all the beliefs are
non-negative and smaller than or equal to unity. This gives us a
full definition of the so-called large polytope LP decoding. One can
run it as is in terms of bit- and check- beliefs, however it may
also be useful to re-formulate the LP procedure solely in terms of
the bit beliefs.

\subsection{Error-floor. Instantons \& Pseudo-Codewords. Pseudo-Codeword Search Algorithm.}

The goal of decoding is to infer the original message from the
received output, ${\bm x}$. Assuming that coding and decoding are
fixed and aiming to characterize the performance of the scheme, one
studies the Frame-Error-Rate (FER)
 $\mbox{ FER} = \int d{\bm x} \,\, \chi_{\mbox{\scriptsize error}}({\bm
   x}) P({\bm x}|{\bm 0})$,
where $\chi_{\mbox{\scriptsize error}} = 1$ if an error is detected
and $\chi_{\mbox{\scriptsize error}} = 0$ otherwise. In a symmetric
channel, FER is invariant with respect to the original codeword,
thus the all-$0$ codeword can be assumed for the input. When {\rm
SNR} is large, FER, as an integral over output configurations, is
approximated by \cite{05SCCV}, $ \mbox{FER}\sim
\sum\limits_{\mbox{\scriptsize inst}} V_{\mbox{\scriptsize inst}}
\times P\left({\bm x}_{\mbox{\scriptsize inst}}|{\bm 0}\right)$,
where ${\bm x}_{\mbox{\scriptsize inst}}$ are the special instanton
configurations of the output maximizing $P({\bm x}|{\bm 0})$ under
the $\chi_{\mbox{\scriptsize error}} = 1$ condition, and
$V_{\mbox{\scriptsize inst}}$ combines combinatorial and
phase-volume factors. Generally, there are many instantons that are
all local maxima of $P({\bm x}|{\bm 0})$ in the noise space. For the
AWGN channel, the instanton estimate for FER at the high {\rm SNR},
$s \gg 1$, is $\sim \exp(-d_{\mbox{\scriptsize inst}} s^2/2)$. In
the instanton-amoeba numerical scheme, suggested in \cite{05SCCV},
instantons with the small effective distances, $d_{\mbox{\scriptsize
inst}}$, were found by a downhill simplex method also called
``amoeba'', with accurately tailored (for better convergence)
annealing. Instantons are closely related to the so-called
pseudo-codewords \cite{95WLK,01FKKR,03Ric,03KV}: decoding applied to
the instanton configuration results in the respective
pseudo-codeword. The effective distance, $d_{\mbox{\scriptsize
inst}}$, characterizing an instanton and its respective
pseudo-codeword, should be compared with the Hamming distance of the
code, $d_{\mbox{\scriptsize ML}}$. Instanton/pseudo-codewords with
$d<d_{\mbox{\scriptsize ML}}$ will completely screen contribution of
the respective codeword to the FER at $s\to\infty$.

In the case of LP decoding, one can actually develop a discrete
computational scheme, coined the Pseudo-Codeword-Search (PCS)
algorithm \cite{06CS}, which allows very efficient calculation of
the low weight pseudo-codewords and the respective instantons. It
was shown in \cite{06CS} that the PCS algorithm converges in a
relatively small number of iterations. The PCS algorithm, repeated
many times picking the initial noise configuration randomly,
generates a set of low-weight pseudo-codewords. Thus,  for the model
$[155,64,20]$ code studied in \cite{06CS,07CSa} some $\sim 200$
pseudo-codewords with the effective weight lower than the Hamming
distance of the code were found.

\subsection{Loop Calculus. Critical Loops. Loop Erasure Algorithm.}

Loop calculus is a technique which allows one to express explicitly
the partition function of the statistical inference problem
associated with Eq.~(\ref{Psx}) in terms of the so-called loop
series \cite{06CCa,06CCb}:
\begin{eqnarray}
 Z&=& Z_0\left(1+\sum\limits_{{\it
C}}r({\it C}) \right),\ \ r({\it C})=\prod\limits_{i,\alpha\in {\it
C}}\mu_\alpha \mu_{i}, \label{z_loop}\\
\mu_{i} &=& \frac{(1-m_i)^{q_i-1}\!+\!(-1)^{q_i}(1+m_i)^{q_i-1}}
 {2(1-m_i^2)^{q_i-1}},\ \ q_i\!=\!\sum\limits_{\alpha\in{\it C}}^{\alpha\ni i}\!1,
\nonumber\\
\mu_{\alpha}\!\!\!\!&=&\!\!\!\!\sum\limits_{\sigma_\alpha}\!b_\alpha(\sigma_\alpha)
 \!\!\prod\limits_{i\in{\it C}}^{i\in\alpha}\!(1\!-\!2\sigma_i\!-\!m_i),\;
m_i\!=\!\sum\limits_{\sigma_i}\! b_i(\sigma_i)(1\!-\!2\sigma_i),
\nonumber
\end{eqnarray}
where $b_\alpha({\bm\sigma}_\alpha)$ and $b_i(\sigma_i)$ are the
beliefs defined on checks and bits according to
Eqs.~(\ref{ba},\ref{bi}) and $Z_0=-\ln F=-\ln(E-S)$ with self-energy
and entropy expressed in terms of the beliefs according to
Eqs.~(\ref{E_Bethe},\ref{S_Bethe}).

The  loop series holds for the BP/MAP relation and it is also well
defined for the LP/ML relation, where transition from the former one
to the later one is according to Eq.~(\ref{rho}). Notice that in the
LP/ML version of the Loop Series $\mu_i$ can be singular, i.e.
$\to\pm\infty$ when $\rho\to\infty$ and $m_i\to\pm 1$, however the
resulting $r({\it C}$ is always finite,  because in this case the
corresponding $\mu_\alpha$ contribution approaches zero. Moreover,
the construction of the Loop Series is such that any individual
$r({\it C})$ is always smaller than unity in absolute value,  both
in the BP and LP cases.

If BP/LP performs well one expects that the loop corrections,
$r({\it C})$, are all significantly smaller than the bare unity.
Failure of the BP/LP decoding signals the importance of some loop
corrections. Even though the number of loops grows exponentially
with the size of the code, not all loops gives comparable
contributions to the loop series. Thus, an important conjecture of
\cite{06CCc} was that for the case of the low-weight
pseudo-codewords (i.e. for the values of log-likelihoods
corresponding to the instanton configuration decoded into the
pseudo-codewords) there exists a relatively simple loop contribution
(or a very few simple contributions), dominating corrections to bare
unity in the loop series. This conjecture was verified in
\cite{06CCc} for the example of the ($\sim 200$) instantons found
for the LP decoding of the $[155,64,20]$ code performing over the
Additive-White-Gaussian-Noise (AWGN) channel. It was demonstrated in
\cite{06CCc} that for each of the instantons, one can indeed
identify the corresponding critical loop, $\Gamma$, giving an
essential contribution to the loop series (\ref{z_loop}) comparable
to the bare LP contribution.

The search for the critical loop suggested in \cite{06CCc} was
heuristic. One searches for a single-connected contribution
associated with a critical loop consisting of checks and bits with
each check connected to only two bits of the loop. According to
Eqs.~(\ref{z_loop}) this contribution to the loop series is the
product of all the triads, $\tilde{\mu}^{(bp)}$, along the loop,
\begin{eqnarray}
 r(\Gamma)\!=\!\prod_{\alpha\in\Gamma}\tilde{\mu}_\alpha,\quad
 \tilde{\mu}_{\alpha}\!=\!\frac{\mu_{\alpha}}
 {\sqrt{(1\!-\! m_i^2)(1\!-\! m_j^2}},
\label{tildemu}
\end{eqnarray}
where for any check $\alpha$ that belongs to $\Gamma$, $i,j$ is the
only pair of $\alpha$ bit neighbors that also belongs to $\Gamma$.
By construction, $|\tilde{\mu}_{\alpha;ij}|\leq 1$. We immediately
find that for the critical loop contribution to be exactly equal to
unity (where unity corresponds to the bare BP term), the critical
loop should consist of triads with all $\tilde{\mu}$ equal to unity
in absolute value. Even if degeneracy is not exact one still
anticipates the contributions from all the triads along the critical
loop to be reasonably large, as an emergence of a single triad with
small $\tilde{\mu}$ will make the entire product negligible in
comparison with the bare BP term. This consideration suggests that
an efficient way to find a single connected critical loop, $\Gamma$,
with large  $|r(\Gamma)|$ consists of, first, ignoring all the
triads with $|\tilde{\mu}|$ below a certain $O(1)$ threshold, say
$0.999$, and, second, checking if one can construct a single
connected loop out of the remaining triads. If no critical loop is
found, we lower the threshold until a leading critical loop emerges.

Applied to the set of instantons of the Tanner $[155,64,20]$ code
with the lowest effective distances this, triad-based search scheme
generates an $r(\Gamma)$ that is exactly unity in absolute value.
This is the special degenerate case in which the critical loop
contribution and the BP/LP contribution are equal to each other in
absolute value. Thus, only the sixth of the first dozen of
instantons has $r(\Gamma)\approx 0.82$ while all others yield
$r(\Gamma)=1$. To extend the triad-based search scheme to the
instantons with larger effective distance, one needs to decrease the
threshold. For the dangerous pseudo-codewords of the $[155,64,20]$
code this always resulted in the emergence of at least one single
connected loop with $r(\Gamma)\sim 1$.

Accounting for a single loop effect (when it is comparable to a bare
(BP) contribution) can be improved through the effective free energy
approach explained in \cite{06CCc}. This approach resulted in the
formulation of renormalized BP equations and the respective
Loop-corrected BP algorithm aimed at solving the renormalized
equations. Modification of the bare BP equations are  well localized
along the critical loop. This observation led to the suggestion an
LP counterpart of the loop-corrected BP, coined the  \underline{\bf
LP-erasure algorithm}:

\noindent $\bullet$ {\bf 1.} Run the LP algorithm. Terminate if LP
succeeds (i.e. a valid code word is found).

\noindent $\bullet$ {\bf 2.} If LP fails,  find the most relevant
loop $\Gamma$ that corresponds to the maximal amplitude $r(\Gamma)$.

\noindent $\bullet$ {\bf 3.} Modify the log-likelihoods
(factor-functions) along the loop $\Gamma$ introducing a shift
towards zero,  i.e. introduce a complete or partial erasure of the
log-likelihoods at the bits. Run LP with modified log-likelihoods.
Terminate if the modified LP succeeds.

\noindent $\bullet$ {\bf 4.} Return to {\bf Step 2} with an improved
selection principle for the critical loop.

This LP-erasure algorithm was tested in \cite{06CCc} on the
$[155,64,20]$ example. The results of the test are remarkable: all
$\sim 200$ low-weight instantons were actually corrected already
with the roughest version of the LP-erasure algorithm, corresponding
to the full erasure of the information (log-likelihoods) along the
critical loop.

\subsection{Facet Guessing Algorithm}

The Facet Guessing (FG) is an improvement of the LP decoder
suggested in \cite{06DW}. This algorithm applies when the bare LP
fails. Failure of LP means that some of the non-strict inequality
constraints in the LP formulation remain inactive for the LP
solution, i.e. the respective strict equalities are not satisfied.
Considering expander codes and the small-polytope version of the LP
decoding, the authors of \cite{06DW} proved that the set of active
constraints of any fractional pseudo-codeword is smaller by a
constant factor than the number of active constraints of any
codeword. This fact was exploited in \cite{06DW} to devise a
decoding algorithm that provably outperforms the LP decoder for
finite blocklengths. The FG algorithm proceeds by guessing the
facets of the polytope, i.e. enforcing the respective inactive
facets to be active with a new equality constraint, and resolving
the linear program on these facets. In its full version, the
algorithm thus consists of the set of modified LP algorithms. The
number of the modified schemes is equal to the number of inactive
facets in the fractional pseudo-codeword solution of the bare LP
algorithm. The configurational output of the FG algorithm is the
output of one modified LP from the set giving the lowest value of
the self-energy (optimization functional). The randomized version of
the FG algorithm consists of picking some fixed fraction of the
modified LP schemes at random from the full set, and then finding
the configuration minimizing the result on the subset. \cite{06DW}
also discussed experimental test of the theory done for couple of
codes, of which one is the $[155,64,20]$ code also considered in
this paper. It was experimentally demonstrated that the randomized
version of the FG algorithm improves the bare LP decoding.

\section{Breaking the critical loop}


We introduce a Bit Guessing (BG) procedure, which is a simplified
version of the FG algorithm.   The simplification comes with a
restriction imposed on the facet-activation (fixing) procedure.  In
BG, one only allows activation (fixing) of the inequalities
associated with bit beliefs, $b_i(\sigma_i)$, and not check beliefs,
$b_\alpha({\bm\sigma}_\alpha)$. Considering values of the
log-likelihoods resulting in a fractional LP pseudo-codeword, one
creates  a set of single bit corrected LP schemes, each different
from the bare LP schemes by only one extra equality condition,
enforcing the value of a bit to be $1$ or $0$. (If the value of the
marginal probability in the bare pseudo-codeword of LP is
fractional, we include two bit-modified LPs in the set,
corresponding to enforcing $0$ and $1$ values for the bit
respectively.  If the marginal probability of a bit is integer we
only includes one bit-modified LP in the set corresponding to fixing
the value of the bit to the integer opposite to the one observed in
the bare pseudo-codeword for the bit.) We run consequently all the
modified LP schemes,  forming the BG set, and choose the result with
the lowest energy functional as the outcome of the bit guessing
procedure.

The FG algorithm was tested on the set of dangerous fractional
pseudo-codewords described in \cite{06CS}. We found that all of the
pseudo-codewords were successfully corrected by FG! In other words,
the output of a corrected LP-scheme with the minimum self-energy is
the right codeword (the all-zero one in the simulations). One also
finds that there always exists a number of successful LP-corrected
schemes, each associated with a different pinned bit. This number
was actually relatively large for the fractional pseudo-codewords
with the lowest weight, $\sim 30-50$, but smaller values  were also
observed for some of the fractional pseudo-codewords with effective
distance from the dangerous range, $[16.4;20]$.

Next for any of the LP-dangerous,  but FG-correctable,
pseudo-codewords, one creates the list of ``successful" bits and
compares this list with the list of bits forming the respective
critical loop found in \cite{06CCc} with the thresholding of the
$\tilde{\mu}$ values. One finds that the set of bits forming the
critical loop forms a relatively small sub-set of the ``successful"
set. In other words,  fixing any bit of the critical loop describing
a dangerous fractional pseudo-codewords leads to correct decoding.

One draws a couple of useful conclusions from this simple
experiment. 1) One finds that the FG algorithm offers a very
successful strategy for decoding in accordance with the main claim
of \cite{06DW}. It corrects all the dangerous pseudo-codewords of
the model $[155,64,20]$ code. 2) The FG correction can be made with
the help of the critical loop procedure of \cite{06CCs}. Finding the
critical loop  helps to reduce the complexity of the operation
beacause it requires adding only one equality constraint to the bare
LP decoding by fixing the value of the marginal probability to zero
or one at any point of the critical loop.

These observations suggest the following decoding algorithm, coined
{\bf Loop Guided Guessing} (LGG):

\noindent $\bullet$ {\bf 1.} Run the LP algorithm. Terminate if LP
succeeds (i.e. a valid code word is found).

\noindent $\bullet$ {\bf 2.} If LP fails,  find the critical loop,
$\Gamma$, the one
 with maximal value of $|r(\Gamma)|$ in the loop series.

\noindent $\bullet$ {\bf 3.} Pick any bit along the critical loop at
random and form two corrected LP schemes,  different from the bare
LP schemes by only one extra equality condition, enforcing the value
of a bit to be $1$ or $0$ respectively.

\noindent $\bullet$ {\bf 4.} Run both LP-corrected schemes and
choose the output which corresponds to the smallest self-energy.
Terminate if the modified LP succeeds.

\noindent $\bullet$ {\bf 5.} Return to {\bf Step 3} selecting
another bit along the critical loop or to {\bf Step 2} for an
improved selection principle for the critical loop if the list of
all the bits along the previously selected loop is exhausted.

Notice that main advantage of the LGG algorithm, in comparison with
the Loop Erasure algorithm of \cite{06CCc}, is in the locality of
the bare algorithm (LP) modification. One finds that breaking a
loop, instead of modifying the algorithm along the loop, is
sufficient for successful decoding.

\begin{figure}
\begin{center}
\includegraphics[width=9.0cm]{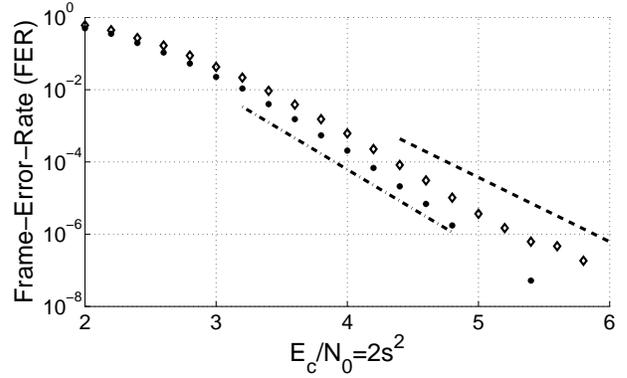}
\end{center}
 \caption{ \label{fig1} Frame Error Rate vs SNR for the $[155,64,20]$ code and the AWGN channel.
 Diamonds represent data of the LP-decoding simulations from
 \cite{07CSa}. Stars stand for the LGG (Loop Guided Guessing)-decoding
 described in the text. Dashed and dashed-dotted lines show
 the MAP and BP/LP decodings $s\to\infty$ asymptotes,
 $\propto \exp(-20*s^2/2)$ and $\propto \exp(-16.407*s^2/2)$ respectively, where
 $20$ is the minimal (Hamming) distance of the code and
 the effective distance of the lowest weight LP pseudo-codeword found for the code is
 $\approx 16.407$.}
\end{figure}

We tested the performance of the LGG algorithm using Monte Carlo
(MC) simulations.  Our starting point was the set of configurations
whose bare LP failed in the MC-LP simulations for the $[155,64,20]$
code discussed in \cite{07CSa}. We apply the LGG algorithm to these
erroneous configurations and observed essential improvement.  Thus
for $s^2=2.4$, only one out of every ten LP-invalid configurations
is not correctable by LGG.   The results of the simulations are
shown in Fig.~(\ref{fig1}). ( Note that performance of the
LP-erasure algorithm of \cite{06CCc} was worse with only a few bare
LP failures corrected.)

For some of these individual configurations, we also run an
exhaustive BG algorithm (checking bit by bit if the respective
bit-corrected LP decodes correctly) and compare the resulting set of
``successful" bits with the set of bits forming the critical  loop.
Very much like the case of the respective dangerous pseudo-codewords
test, we found strong correlations between the two sets: bits of the
critical loop typically belong to the ``successful" set. This
justifies our decision to select the special bit along the critical
loop at random,  thus supporting the conjecture that a pin-point
bit-local correction of LP is sufficient for breaking the loop and
successful decoding.

Notice that configurations accessable at SNRs from Fig.~\ref{fig1}
via MC simulations  are typically those with relatively large
effective distances, $\sim 30-40$, while the Hamming distance of the
code is $20$ and the effective distance of the most dangerous
pseudo-codeword of the bare LP is $\approx 16.4$. We expect,
however, that majority of these configurations are from valleys of
the Bethe Free Energy functional with local minima correspondent to
effective distances from the $[16.4;20]$ range. See \cite{05SC} for
a related discussion of why the FER asymptote at moderate SNR shows
behavior controlled by pseudo-codewords with much smaller effective
distance than those representing a given SNR.

\section{Path forward}

Let us conclude listing some future problems/challenges:

\noindent $\bullet$ The LGG algorithm should be tested on longer
practically relevant codes.

\noindent $\bullet$ The LGG performance can be improved if a better
algorithm for finding the critical loop is implemented. An LDPC code
can be replaced by its MAP-equivalent dendro-counterpart
\cite{07CSa}. Then,  the problem of finding the single-connected
loop with the largest value of $|r(C)|$ is reduced to finding the
shortest path on the undirected graph with possibly negative
weights, however with a guarantee that all loop contributions over
the graph are positive. One may hope to develop an efficient graph
algorithm for solving this problem.

\noindent $\bullet$ Modern schemes of LDPC ensemble optimization
\cite{06AMU} are very successful in dealing with the water-fall
domain,   where performances of almost  all codes from the given
ensemble are identical.   It is known however that different codes
from the same ensemble show big performance variations in the
error-floor domain if a standard,  not yet optimized, decoding is
utilized. This problem is a serious handicap for the successful use
of random LDPC codes in the demanding high SNR regime.  We plan to
apply the decoding improvement strategy to the optimized LPDC
ensembles,  with the hope that the algorithm improvement may be
capable in lowering the error-floor for a majority of codes from the
ensemble.

The author acknowledge very useful, inspiring and fruitful
discussions with V. Chernyak, A. Dimakis, M. Stepanov and M.
Wainwright. This work was carried out under the auspices of the
National Nuclear Security Administration of the U.S. Department of
Energy at Los Alamos National Laboratory under Contract No.
DE-AC52-06NA25396.

\end{document}